\documentclass[12pt]{article}
\usepackage{pstricks}
\usepackage{pst-coil}
\usepackage{amssymb}
\newcommand{\be}{\begin{eqnarray}}
\newcommand{\ee}{\end{eqnarray}}
\newcommand{\mh}{\hat{{\cal M}}}
\newcommand{\gh}{\hat {g}}
\newcommand{\vh}{\hat {v}}
\newcommand{\gamh}{\hat{\gamma}}
\newcommand{\etah}{\hat{\eta}}
\newcommand{\Rh}{\hat{R}}
\newcommand{\Dh}{\hat{D}}
\newcommand{\nbh}{\hat{\nabla}}
\newcommand{\dto}{{\dot{t}}_0}
\newcommand{\ddto}{{\ddot{t}}_0}

\newtheorem{lem}{Lemma}
\begin{document}
\title{A note on behaviour at an isotropic singularity}
\author{Brien C. Nolan\\School of Mathematical Sciences,\\
Dublin City University,\\Glasnevin, Dublin 9\\Ireland}
\date{\today}
\maketitle
\begin{abstract}
The behaviour of Jacobi fields along a time-like geodesic running into an isotropic singularity is studied. It is shown that the Jacobi fields are crushed to zero length at a rate which is the same in every direction orthogonal to the geodesic. We show by means of a counter-example that this crushing effect depends crucially on a technicality of the definition of isotropic singularities, and not just on the uniform degeneracy of the metric at the singularity.
\\PACS: 04.20.Dw, 98.80.H, 04.20.-q
\end{abstract}

\section{Introduction}
Isotropic singularities were introduced into the literature in \cite{gw85} as a tool with which to study Penrose's Weyl Tensor Hypothesis and in an attempt to unify various studies of the initial cosmological singularity. Since then, they have received a lot of attention in the context of this and other hypotheses; in particular, the Vanishing Weyl Tensor Conjecture (that a cosmology admitting an isotropic singularity with initially vanishing Weyl tensor evolves uniquely to a Robertson-Walker universe) has recently been proven for a class of equations of state \cite{ta}. The basic idea of an isotropic singularity is that the singular behaviour may be removed by a conformal rescaling, where the conformal factor vanishes at the singularity. The formal definition runs as follows \cite{gw85,se98}.

A space-time $(\mh,\gh)$ is said to admit an {\em isotropic singularity} if there exists a space-time $({\cal M},g)$, a smooth cosmic time function $t$ defined on ${\cal M}$ and a function $\Omega(t)$ such that
\newcounter{conds}
\begin{list}
{(\roman{conds})}{\usecounter{conds}}
\item
$\mh$ is the open submanifold $t>0$.
\item
$\gh_{ab}=\Omega^2(t)g_{ab}$ on $\mh$, with $g_{ab}$ $C^3$ and non-degenerate on an open neighbourhood of $t=0$.
\item
$\Omega\in C[0,b]\cap C^3(0,b]$ for some $b>0$, $\Omega(0)=0$ and $\Omega(t)>0$ for $t\in(0,b]$.
\item
The limit
\[\lambda =\lim_{t\to 0^+}\frac{\Omega\Omega^{\prime\prime}}{(\Omega^\prime)^2}\neq 1\]
exists, where the prime indicates differentiation with respect to $t$.
\end{list}
The low differentiability requirement on $\Omega$ at $t=0$,
which is identified with the singularity, is required to allow
sufficient generality amongst this class of space-times. (If $\Omega\in C^\infty[0,b]$ and space-time is filled with a perfect fluid with a barotropic equation of state, then the fluid pressure $p$ and density $\rho$ must obey $\rho = 3p$ \cite{newman}.)
The following results may be derived using the conditions imposed upon $\Omega$ and are essential for what follows \cite{scott}:
\begin{lem}
$\lambda < 1$.
\end{lem}
\begin{lem}
$\lim_{t\to 0^+} (\Omega^\prime)^{-1}\Omega=0.$
\end{lem}
\begin{lem}
$ (\Omega^\prime)^{-1}\Omega \sim (1-\lambda)t$ as $t\to 0^+$.
\end{lem}

Such singularities are usually studied in a cosmological context, and so it is generally assumed that there exists a congruence of time-like curves in $(\mh,\gh)$ representing the aggregate flow of matter in the space-time. We will assume this congruence to be geodesic. So let $\vh^a$ be the unit tangent to a congruence of time-like geodesics of $(\mh,\gh)$, with typical member $\gamh$.
We note that as the definition of an isotropic singularity is independent of the matter content of the space-time in question, this congruence need not have the interpretation mentioned above. So we are studying any time-like geodesic (subject to one further restriction given below) which emanates from an isotropic singularity.
We can associate with this congruence a congruence of time-like curves in $({\cal M},g)$ with unit tangent $v^a=\Omega \vh^a$. The congruence associated with $\vh^a$ is said to be {\em regular} if the conformal image $v^a$ is $C^3$ on an open neighbourhood of $t=0$.
So let $(\mh,\gh)$ admit an isotropic singularity, and a regular time-like geodesic congruence with unit tangent $\vh^a$.

Our aim here is to study carefully the idea of the isotropy of the singularity. To this end, we show that {\em every} Jacobi field along $\gamh$ is crushed to zero length {\em at the same rate} as the singularity is approached. Thus physical objects carried along $\gamh$ are crushed to zero volume at the same rate in every direction; the crushing effect is isotropic. The rate at which this crushing takes place depends only on the index $\lambda$. This is perhaps not surprising; indeed it verifies on a rigorous level part of what one wants to model for such singularities (see Tod's comments about blobs of fluid in \cite{tod}), but we shall see that the crushing effect arises not just from the fact that $\Omega$ vanishes at the singularity, but that the regularity condition of the existence of the limit $\lambda$ and the assumption that $t$ is smooth on $({\cal M},g_{ab})$ are crucial. As a useful by-product, we obtain the general asymptotic behaviour of the conformal factor $\Omega$ as a function of proper time along $\gamh$ in the limit as the singularity is approached. We use the metric and curvature conventions of \cite{he}.

\section{Decomposition and solution of the geodesic deviation equation.}
A Jacobi field along $\gamh$ is a vector field $\etah^a$ orthogonal to $\vh^a$ which satisfies the geodesic deviation equation
\begin{equation}
\Dh^2\etah^a = -{\Rh^a}_{\,\,\,bcd}\etah^c\vh^b\vh^d, \label{eq3}
\end{equation}
where ${\Rh^a}_{\,\,\,bcd}$ is the Riemann tensor of $(\mh,\gh)$ and $\Dh=\vh^a\nbh_a$ is the covariant derivative along $\gamh$. The solutions of (\ref{eq3}) form a six-dimensional vector space orthogonal to $\vh^a$.
To solve (\ref{eq3}), we use the assumptions above to identify the singular and non-singular terms of the right hand side. We obtain the asymptotic behaviour of these terms in the limit $s\to 0^+$, where $s$ is the proper time along an integral curve of $\vh^a$, whose origin is fixed so that the (past) singularity is approached as $s\to 0^+$.

By the standard formulae for conformally related metrics \cite{he},
\begin{eqnarray}
{\Rh^a}_{\,\,\,bcd}\etah^c\vh^b\vh^d&=&
\Omega^{-2}{R^a}_{bcd}\etah^c v^b v^d \nonumber\\
&&+\frac14 (\delta^a_c \Omega_{bd}-g_{bc}\Omega^a_d-\delta^a_d \Omega_{bc}+g_{bd}\Omega^a_c)\etah^c v^bv^d,
\label{eq4}
\end{eqnarray}
where
\[ \Omega_{ab}=-4\Omega^{-3}\nabla_a\nabla_b\Omega+8\Omega^{-4}\nabla_a\Omega\nabla_b\Omega-2\Omega^{-4}\nabla_c\Omega\nabla^c\Omega g_{ab}.\]

The assumptions made above allow us to simplify the right hand side of
(\ref{eq4}) considerably. We note first that as $\vh^a$ is assumed to be geodesic, then the relationship between the acceleration vectors of $\vh^a$ and $v^a$ which comes from the conformal transformation
\[ \Dh\vh^a=\Omega^{-2}Dv^a +\Omega^{-3}D\Omega v^a +\Omega^{-3}\nabla^a\Omega\]
yields
\begin{equation}
\nabla^a\Omega = -D\Omega v^a-\Omega a^a,\label{eq5}
\end{equation}
where $D=v_a\nabla^a$ and $a^b=v^a\nabla_av^b$. Then also
\[ \nabla_a\Omega\nabla^a\Omega = -(D\Omega)^2+\Omega^2a_ba^b\] and
\begin{equation}
\etah_a \nabla^a\Omega =-\Omega a_b\etah^b,
\label{eq6}
\end{equation}
and by definitions, $\etah_a \vh^a = \etah_a v^a =0$.

The scalar term in the first bracketed term of (\ref{eq4}) can be written
\begin{equation}
\Omega_{bd}v^bv^d = -4\Omega^{-1}\Dh^2\Omega +2\Omega^{-2}(\Dh\Omega)^2-2\Omega^{-2}a_ba^b,\label{eq7}
\end{equation}
where we have used $\Dh =\Omega^{-1}D$ when acting on scalars.

Next, a careful calculation relying heavily on (\ref{eq5}) and geodesy of $\vh^a$ shows that
\begin{eqnarray}
\Omega_{bc}\etah^cv^b&=&4\Omega^{-2}Da_b\etah^b.\label{eq8}
\end{eqnarray}
Finally, the same conditions show that
\begin{eqnarray}
\Omega^a_b\etah^b&=&
4\Omega^{-2}(\Dh\Omega\nabla^av_b+(\Dh\Omega v^a+a^a)a_b+\nabla^aa_b)\etah^b\nonumber\\
&&+2\Omega^{-2}((\Dh\Omega)^2-a_ba^b)\etah^a.
\label{eq9}
\end{eqnarray}

Thus from equation (\ref{eq4}) and equations (\ref{eq6})-(\ref{eq9}) the geodesic deviation equation (\ref{eq3}) reads
\begin{equation}
\Dh^2\etah^a+{\Lambda^a}_b\etah^b =0,\label{eq10}
\end{equation}
where
\[
{\Lambda^a}_b
={{\Lambda_1}^a}_b+{{\Lambda_2}^a}_b+{{\Lambda_3}^a}_b,
\]
with
\begin{eqnarray}
{{\Lambda_1}^a}_b&=&\Omega^{-2}({R^a}_{cbd}v^cv^d-a^aa_b-v^aDa_b-\nabla^aa_b)=:\Omega^{-2}U^a_b,\label{leq1}\\
{{\Lambda_2}^a}_b&=&-\Omega^{-2}\Dh\Omega(\nabla^av_b+v^aa_b)=:
\Omega^{-2}\Dh\Omega W^a_b,\label{leq2}\\
{{\Lambda_3}^a}_b&=&-\Omega^{-1}\Dh^2\Omega\delta^a_b.
\label{leq3}
\end{eqnarray}
We note an essential point for what follows: the definition of an isotropic singularity and the assumption that $\vh^a$ is regular, i.e. that $v^a$ is $C^3$, implies that the tensors $U^a_b, W^a_b$ are {\em finite} at the singularity.
In the usual way, we can project (\ref{eq10}) onto a parallel propagated tetrad, and so treat $\etah^a$ as a Euclidean 3-vector and $\Dh$ as an ordinary derivative (see section 4.1 of \cite{he}).

In order to obtain the asymptotic behaviour of solutions of (\ref{eq10}) as $s\to 0^+$, we must determine the asymptotic behaviour of $\Omega$ and its $s-$derivatives ($\Dh\Omega$ etc.) in this limit.
As a first step towards this, note that
\[ \Dh\Omega =\Omega^{-1}D\Omega =\Omega^{-1}\Omega^\prime Dt.\]
By the assumption that $t$ is a smooth cosmic time function throughout a neighbourhood of $t=0$ in $({\cal M},g)$ and the fact that $v^a$ is $C^3$, we must have that\footnote{It is of course possible that $Dt|_{t=0}=0$. However the following analysis still holds in this case, as regardless of whether or not $\dto^{-1}\ddto$ converges or diverges, the dominant term in ${L^a}_b$ is always ${{L_3}^a}_b$ as can be seen from equations (\ref{eq19}-\ref{eq21}) in the text.}
\[ \dto:=\lim_{s\to 0^+}Dt >0.\]
(Recall that $s\to0^+$ corresponds to $t\to 0^+$.) Thus
\begin{equation}
\Dh\Omega \sim \dto\frac{\Omega^\prime}{\Omega}.\label{eq13}
\end{equation}
Similarly, the limit
\[ \ddto:=\lim_{s\to 0^+}D^2t \]
must exist and be finite, but may be positive, negative or zero.
From these observations, we find that
\begin{equation}
\Dh^2\Omega \sim \Omega^{-2}\Omega^\prime\ddto +(\Omega^{-2}\Omega^{\prime\prime}-\Omega^{-3}{\Omega^\prime}^2)\dto^2.
\label{eq14}
\end{equation}

Next, consider the term $\frac{\Omega^2/\Omega^\prime}{s}$. Both numerator and denominator vanish in the limit $s\to 0$ (see Lemma 2 above), and so we can try to use l'Hopital's rule to evaluate the limit of the ratio. This gives a finite and hence legitimate result:
\begin{eqnarray}
\lim_{s\to 0^+}\frac{\Omega^2/\Omega^\prime}{s}
&=&\lim_{s\to 0^+}\left(2\Omega-\frac{\Omega^2\Omega^{\prime\prime}}{(\Omega^\prime)^2}\right) \Dh t\nonumber\\
&=&\lim_{s\to 0^+}\left(2-\frac{\Omega\Omega^{\prime\prime}}{(\Omega^\prime)^2}\right)Dt\nonumber\\
&=&\dto(2-\lambda).\label{eq15}
\end{eqnarray}
The same reasoning may be applied to the quantity $\ln\Omega/\ln s$:
\begin{eqnarray}
\lim_{s\to 0^+}\frac{\ln\Omega}{\ln s}&=&\lim_{s\to 0^+}\frac{\Dh(\ln\Omega)}{\Dh(\ln s)}\nonumber\\
&=&\lim_{s\to 0^+}\frac{s\Omega^\prime Dt}{\Omega^2}\nonumber\\
&=&\frac{1}{2-\lambda},
\label{eq16}
\end{eqnarray}
where we have used (\ref{eq15}). This yields
\begin{equation}
\Omega \sim s^{\frac{1}{2-\lambda}}.\label{eq17}
\end{equation}
Recall that $\lambda<1$, so that this power is always positive. We note that the relations (\ref{eq15}) and (\ref{eq17}) hold for {\em all} isotropic singularities with a regular congruence of time-like curves: the assumption that $\gamh$ is geodesic was not used in the derivation of these equations.

From equations (\ref{eq13})-(\ref{eq17}) we can determine the asymptotic behaviour of the singular terms in (\ref{eq10}). It is convenient to rewrite (\ref{eq10}) in the form
\begin{equation}
s^2\Dh^2\etah^a +{L^a}_b\etah^b=0,
\label{eq18}
\end{equation}
with the obvious defintion of ${L^a}_b$ (and its constiuents ${{L_1}^a}_b$ etc. below).
The singular coefficient of ${{L_1}^a}_b$ is
\begin{equation}
s^2\Omega^{-2} \sim s^{\frac{2-2\lambda}{2-\lambda}}.
\label{eq19}
\end{equation}
We note that
\[ 0<\frac{2-2\lambda}{2-\lambda}<2.\]
The singular coefficient of ${{L_2}^a}_b$ is
\begin{eqnarray}
-s^2\Omega^{-2}\Dh\Omega&=&-s^2\Omega^{-3}\Omega^\prime Dt\nonumber\\
&\sim&-\frac{1}{2-\lambda}s^{\frac{1-\lambda}{2-\lambda}}.
\label{eq20}
\end{eqnarray}
Notice that this term dominates the coefficient of ${{L_1}^a}_b$.
The  coefficient of ${{L_3}^a}_b$ satisifes
\begin{eqnarray}
-s^2\Omega^{-1}\Dh^2\Omega
&\sim&
-s^2\Omega^{-3}\Omega^\prime\ddto-s^2
(\Omega^{-3}\Omega^{\prime\prime}-\Omega^{-4}(\Omega^\prime)^2)\dto^2\nonumber\\
&=&
-s^2\Omega^{-3}\Omega^\prime\ddto-\frac{s^2(\Omega^\prime)^2}{\Omega^4}
\left(\frac{\Omega\Omega^{\prime\prime}}{(\Omega^\prime)^2}-1\right)
\dto^2\nonumber\\
&\sim&
\frac{1-\lambda}{(2-\lambda)^2}-\frac{\ddto}{(2-\lambda)\dto}
s^{\frac{1-\lambda}{2-\lambda}}.\label{eq21}
\end{eqnarray}
Thus the dominant term in ${L^a}_b$ is ${{L_3}^a}_b$, and we can write
\[ {L^a}_b=\frac{1-\lambda}{(2-\lambda)^2}\delta^a_b +
s^{\frac{1-\lambda}{2-\lambda}}{M^a}_b,\]
where ${M^a}_b(s)$ is continuous at $s=0^+$. The $s-$dependence
of the second term here indicates that ${L^a}_b$ has a low degree
of differentiability, indicating that the solutions of (\ref{eq18})
cannot be written as power series in $s$. Nevertheless, the leading
order behaviour of a fundamental matrix for (\ref{eq18}) can be found
by looking for a formal solution whose entries are formal logarithmic
sums, that is functions of the form
\[ f=\sum_{j,k=0}^\infty f_{jk}s^{\mu_j}(\log s)^k,\]
where the $f_{jk}$ are formal Laurent series in $s$ which vanish for
sufficiently large $j+k$ and the $\mu_j$ are complex numbers \cite{codlev}.
It is easily verified that the leading order term in such a series
corresponds to a fundamental matrix for the diagonal equation
\[ s^2\Dh^2\etah^a +\frac{1-\lambda}{(2-\lambda)^2}\etah^a =0.\]
A fundamental matrix for this equation is given by
\[ y(s)I_3,\]
where $I_3$ is the $3\times3$ identity matrix and $y(s)$ satisfies the
scalar equation
\begin{equation}
 s^2y^{\prime\prime}(s)+ \frac{1-\lambda}{(2-\lambda)^2}y(s)=0.\label{eq22}
\end{equation}
The general solution of this equation is easily found using the method of
Frobenius. There are slight variations (the presence of logarithmic terms
in the solution) depending on whether or not $\lambda =2n/(n+1)$ for some
integer $n$; however all that concerns us is that the general solution
satisfies
\[ y(s) \sim c s^\nu,\]
where $c$ is constant and $\nu$ (positive)
is the smaller of $1/(2-\lambda)$ and
$(1-\lambda)/(2-\lambda)$. Thus the general solution of (\ref{eq22})
satisfies $y\to 0$ as $s\to 0^+$.

Hence there exists a fundamental matrix $E$ for (\ref{eq18}) which
satisfies
\[ E(s)\sim y(s)I_3,\qquad {\mbox{as }} s\to 0^+\]
and $y\to 0$ as $s\to 0^+$. Therefore every solution
$\etah^a$ of (\ref{eq18}) shrinks to zero as the singularity
is approached; the crushing effect of the singularity is the same
in every direction. This reinforces the notion of isotropic behaviour
at such singularities. We note further that the rate at which the Jacobi
fields are crushed depends only on the index $\lambda$ and not on any
details of the region of space-time from which the singularity is being
approached. Thus as one would expect, there is an accompanying degree of
homogeneity allied to the isotropic behaviour at the singularity.

\section{Unexpected behaviour at a singularity which is nearly isotropic}
To further elucidate the role played by the defining conditions of an
isotropic singularity, we consider an example of a space-time which has
an initial singularity with the same broad features as an isotropic
singularity, but which does not qualify as such on a technical basis.
We show that the approach to this singularity is not accompanied by the
isotropic crushing found above for isotropic singularities.

We consider the spherically symmetric conformally flat space-time
(so that $g_{ab}$ above is the metric tensor $\eta_{ab}$ of Minkowski space-time)
with conformal factor
\begin{equation}
\Omega(r,T)=T\exp\left(-\frac{r^2}{T^n}\right),\label{eq23}
\end{equation}
where $n>0$ and $T,r$ are respectively a global Cartesian time coordinate and the radius function of Minkowski space-time. We take $T>0$; then $T=0$ is a past singularity of the space-time.
Note that since $\frac{r^2}{T^n}\geq 0$, $\Omega\to 0$ as $T\to0^+$ for all values of $r$. For $T>0$, $r=0$ is the regular centre of the space-time, and is a time-like geodesic $\gamh$ with unit tangent
\[ \vh^a=\Omega^{-1}\delta^a_T.\]
The equations governing the Jacobi fields along such a geodesic are given in \cite{bn99}. The two tangential Jacobi fields along the radial geodesic $r=r_0=$ constant have norms of the form
\begin{equation}
x(s)=c_1R(s)\int \frac{ds}{R^2(s)} + c_2R(s),\label{eq24}
\end{equation}
where $R$ is the radius function of the space-time, $s$ is proper time along the geodesic and $c_{1,2}$ are arbitrary constants giving the 2-parameter general solution.
Along the geodesic in question, $R=r_0\Omega(r_0,T)$ and $ds=\Omega dT$. Fixing the origin of $s$ so that the singularity $T=0$ occurs at $s=0$, we see that, modulo an irrelevant multiplicative constant,
\[ \lim_{s\to 0^+} x(s) =\lim_{T\to 0^+} \Omega(r_0,T)\int \frac{dT}{\Omega(r_0,T)}.\]
Using l'Hopital's rule, this gives
\[ \lim_{s\to 0^+} x(s) =-\lim_{T\to 0^+}\frac{T^{n+1}}{T^n+nr_0^2} = 0.\]
Notice that this result applies in the case $r_0=0$, and so applies to the geodesic $\gamh$.

The norm $y(s)$ of the radial Jacobi field is governed by the equation
\begin{equation}
\Dh^2 y(s)+F y(s)=0,\label{eq25}
\end{equation}
where
\[ F(r,T) = \Omega^{-4}(\Omega_T^2-\Omega_r^2)-\Omega^{-3}(\Omega_{TT}-\Omega_{rr}),\]
where the subscripts indicate partial derivatives.
Using $ds=\Omega dT$, we have
\[ \Dh^2 y =\Omega^{-2}{\ddot y}-\Omega^{-3}\Omega_T{\dot y},\]
where the overdot represents differentiation with respect to $T$.
Then a straightforward calculation shows that, along $\gamh$, equation (\ref{eq25}) may be written
\[ T^2{\ddot y} -T{\dot y} - (2T^{2-n}-1)y=0.\]
If $n<2$, the general solution of this equation vanishes at $T=0$. We consider the case $n=2$. Then the general solution may be written down explicitly;
\[ y(t) =c_1T^{1+\sqrt{2}}+c_2T^{1-\sqrt{2}}.\]
Thus the general solution, and hence the general radial Jacobi field, {\em blows up} at the singularity $T=0$. The same holds for $n>2$. So the deformation along this geodesic is {\em not} isotropic; infinite stretching occurs along the radial direction, and infinite crushing along the tangential directions. Note however that the characteristic volume along $\gamh$ that has norm $V=|x^2y|$ satisfies
\[ V\sim V_0 T^{3-\sqrt{2}}\to 0.\]

We investigate which of the conditions for an isotropic singularity this example fails to satisfy. We fix $n=2$. Then
\be \eta^{ab}\Omega_a\Omega_b = -\exp\left(-\frac{2r^2}{T^2}\right)(1+4\frac{r^4}{T^4}).\label{gradom}\ee
We note therefore that $\nabla^a\Omega$ is timelike for all $(r,T)$ with $T\neq 0$, but is {\em degenerate} for $\Omega=0$. Thus $\Omega=0$ does not represent a hypersurface within the family of hypersurfaces $\Omega=$constant. However we know that $\Omega=0\Leftrightarrow T=0$, so the singularity may be identified with the space-like hypersurface $T=0$. (This is further verified by showing that each point $(T=0,r=r_r)$ of the hypersurface is a past endpoint for both an ingoing and an outgoing radial null geodesic of the space-time.)

According to the definition of \cite{newman}, the vanishing of $\nabla^a\Omega$ on $\Omega=0$ disqualifies this example from membership of the class of isotropic singularities. Similarly, it is not possible to find an appropriate time function $Z=Z(\Omega)$ satisfying $Z(0)=0$ and $\nabla^aZ\neq 0$. That is, in the language of \cite{ta}, there is no conformal gauge choice which allows this example to be considered an isotropic singularity (see also \cite{tod}). This failure to qualify may be clarified by considering how the example fails to qualify on the basis of the definition above, which does not {\em explicitly} include the requirement $\nabla^a\Omega\neq 0$.

First, we note that taking $t=\Omega$, the function $t$ is a cosmic time function on ${\cal{M}}$, i.e. $t\in C^0({\cal{M}})$ and increases along every future-directed causal curve of ${\cal{M}}$\footnote{This is the definition of \cite{BE}. Note the inequivalence to the definition that $\nabla^at$ is time-like everywhere. The former definition is more general, allowing (i) less smooth functions; (ii) examples such as $t=T^3$, where $T$ is a global Euclidean time co-ordinate on Minkowski space-time.}. This is clear for the open submanifolds $t\neq 0$, as equation (\ref{gradom}) shows. Let $s^a$ be tangent to a future-pointing causal curve $\sigma$ of ${\cal{M}}$. Then $s^a\nabla_at>0$ for all $t\neq 0$, but $s^a\nabla_at|_{t=0}=0$, so certainly $t$ is non-decreasing everywhere along $\sigma$. But crossing $t=0$ is equivalent to crossing $T=0$; this increases from negative to positive along $\sigma$, which is future directed. Hence $t$ also increases from negative to positive along $\sigma$. So $t$ does indeed increase, in the strict sense, along every future-directed causal curve of ${\cal{M}}$; $t|_{\sigma(\tau_2)}>t|_{\sigma(\tau_1)}$ for values $\tau_2>\tau_1$ of a parameter $\tau$ along $\sigma$ which increases into the future. In particular, we have $Dt=1$ along the geodesic $r=0$, and so $\dto=1$ for this example.

With $t=\Omega$, conditions (i)-(iv) above are trivially satisfied, with $\lambda=0$. Also, the geodesic $\gamh$ is regular with geodesic conformal image. Thus it is only the failure of $t:=\Omega(r,T)$ to be a {\em smooth} cosmic time function on $({\cal M},\eta_{ab})$ which precludes this example from satisfying all the hypotheses of an isotropic singularity. In fact $t$ fails to be smooth ($C^\infty$)  only at the point $(r,T)=(0,0)$. Therefore this example shows the importance of this assumption for isotropic singularities. In terms of the relationship with the definitions of \cite{newman} and \cite{ta}, as pointed out by Tod in \cite{tod}, the current definition (\cite{gw85,se98}) assumes implicitly that among the allowed cosmic time functions of the definition above, one - ${\bar t}$ say - can be chosen so that $\nabla^a{\bar{t}} \neq 0$ at the singularity. Assuming smoothness for a particular choice $t$ would allow one to find the ``correct'' ${\bar{t}}={\bar{t}}(t)$ (this issue may deserve closer attention). The lack of smoothness of our $t$ in this example indicates that no such ${\bar{t}}$ exists; the degeneracy of $\Omega$ at the singularity, i.e. $\nabla^a\Omega = 0$ is gauge-invariant.

\section{Conclusions}
We have studied some technicalities of the definition of isotropic
singularities. As one would expect for such a definition, the behaviour
at the singularity is indeed isotropic, in the sense that Jacobi fields
carried along a geodesic running into such a singularity are crushed
uniformly in every direction. This verifies in a rigorous manner what
one intends to model with such a definition; a blob of matter would be
crushed uniformly to zero volume as the singularity is approached.
We have also found homogeneous behaviour at the singularity; the rate
at which the Jacobi fields are crushed depends only on the index $\lambda$
of the definition above, and not on the region of space-time from which
the singularity is approached. Thus although space-times containing
isotropic singularities are usually inhomogeneous \cite{se98},
the inhomogeneity is washed out in the approach to the singularity.

We have derived this result for time-like geodesics subject to a certain regularity condition. For non-geodesic motions, the deviation equation becomes more complicated (see equation (4.8) of \cite{he}). However the generalisation of the result to bounded acceleration curves (for example the fluid flow lines in the examples of \cite{mars}) does not appear to be inconceivably difficult: since in this case equation (\ref{eq5}) remains true in the limit $\Omega\to 0$, the asymptotic analysis should follow through.

The example of section 3 above shows how these conclusions depend
crucially on a technicality of the definition of \cite{gw85,se98}, namely that the conformal
factor is a function of a cosmic time function which is {\em smooth}
on the background space-time. In our example, the cosmic time function fails to be even $C^1$ throughout the hypersurface $t=0$. 
This manifests itself in the fact that regardless of how one tries to choose the conformal factor $\Omega$ and appropriate time function $Z$ (see \cite{tod}) for this particular space-time, one finds the degeneracy $\nabla^aZ=0$ at the singularity. 

\section*{Acknowledgement}
I am grateful to an anonymous referee for constructive criticism.

\end{document}